\documentclass[a4paper,11pt]{article}
\usepackage{pos}
\usepackage{float}
\usepackage{lineno}
\usepackage{graphicx}

\newcommand*{\mkblue}[1]{{\color{black}{#1}}}
\newcommand*{\mkred}[1]{{\color{black}{#1}}}
\newcommand*{\mkgreen}[1]{{\color{black}{#1}}}
\title{First look at data from the 13-antenna setup of GRANDProto300 in northwest China}
 \ShortTitle{First data of GP13}

\author*[a]{Peng-Xiong Ma}
\author[a,b]{Bo-Hao Duan}
\author[c]{Xin Xu}
\author[a,b]{Ke-Wen Zhang}
\author[a]{Kai-Kai Duan}
\author[a]{Shen Wang}
\author[a,b]{Yi Zhang}
\author[c]{Peng-Fei Zhang}
\onbehalf{for the GRAND Collaboration}

\affiliation[a]{Key Laboratory of Dark Matter and Space Astronomy, Purple Mountain Observatory, Chinese Academy of Sciences\\
  No. 10 Yuanhua Road, Nanjing, China}

\affiliation[b]{School of Astronomy and Space Science, University of Science Technology of China\\
No. 96 Jinzhai Road, Hefei , China}

\affiliation[c]{School of Electronic Engineering,Xidian University\\
No. 2 South Taibai Road, Xi'an, China}

\emailAdd{mapx@pmo.ac.cn}
\abstract{The Giant Radio Array for Neutrino Detection (GRAND) is an envisioned observatory of ultra-high-energy neutrinos, cosmic rays, and gamma rays, with energies above 100 PeV. GRAND targets the radio signals emitted by extensive air showers induced by the interaction of ultra-high-energy particles in the atmosphere, using an array of 200,000 radio antennas split into sub-arrays deployed worldwide. \mkred{GRANDProto13 (GP13) is a 13-antenna demonstrator array deployed in February 2023 in the Gansu province of China, as a precursor for GRANDProto300, which will validate the detection principle of the GRAND experiment.} Its goal is to measure the radio background present at the site, validate the design of the detection units \mkgreen{and} develop \mkred{an autonomous radio trigger for air showers}. We will describe GP13 and its operation, and show preliminary results on noise monitoring. 

}

\ConferenceLogo{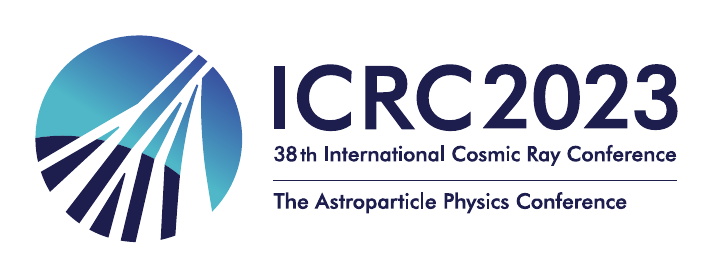}

\FullConference{%
38th International Cosmic Ray Conference (ICRC2023)\\
  26 July - 3 August, 2023\\
  Nagoya, Japan}

\begin{document}
\maketitle

\section{Introduction}

The discovery of cosmic rays over a century ago marked a significant milestone in scientific exploration. In the past few decades, substantial advancements have been made in this field. However, despite these achievements, there remain several fundamental questions that demand answers. Specifically, the origin of \mkgreen{ultra-high-energy cosmic rays} and the mechanisms behind their acceleration to such extraordinary energies continue to elude researchers.

In recent years, a novel approach utilizing radio radiation emitted during the interaction of \mkgreen{ultra-high-energy 
cosmic rays} with the Earth's atmosphere has emerged as a promising avenue for cosmic-ray measurement \cite{Ardouin:2005qe,Ardouin:2009zp,deVries:2015oda,Ardouin:2010gz,Charrier:2018fle}. Pioneering experiments and simulation studies have demonstrated the efficacy and efficiency of this method in energy reconstruction and the differentiation of \mkgreen{cosmic-ray} chemical compositions \cite{LOPES:2005ipv,Huege:2016veh,2016Natur.531...70B,PierreAuger:2016vya,ANITA:2018vwl,2023arXiv230309249C}.

\section{GRANDProto13}

\subsection{Design}
\label{sec:design}

The GRAND project represents a vast and ambitious international collaboration aimed at elucidating the origins of high-energy cosmic rays and discerning their chemical composition \cite{GRAND:2018iaj}. A staged approach was chosen to accomplish this endeavor. A key step is GRANDProto300 \cite{Duan:2023tdh,Martineau-Huynh:2019bgk}, a \mkred{300-antenna} prototype array deployed over 200~$\mathrm{km^2}$ in the Gobi desert \mkred{in China}, which is presently being built. Extensive site surveys have been conducted since 2019, ultimately leading to the selection of the XiaoDuShan site (E $93.94177^\circ$, N $40.99434^\circ$), situated approximately 150 km north of Dunhuang city. This site exhibits \mkred{significantly reduced} levels of radio background noise across a wide frequency range, ranging from \mkblue{several} MHz to hundreds of MHz. 

\begin{figure*}[]
\centering
\includegraphics[width=7.9cm]{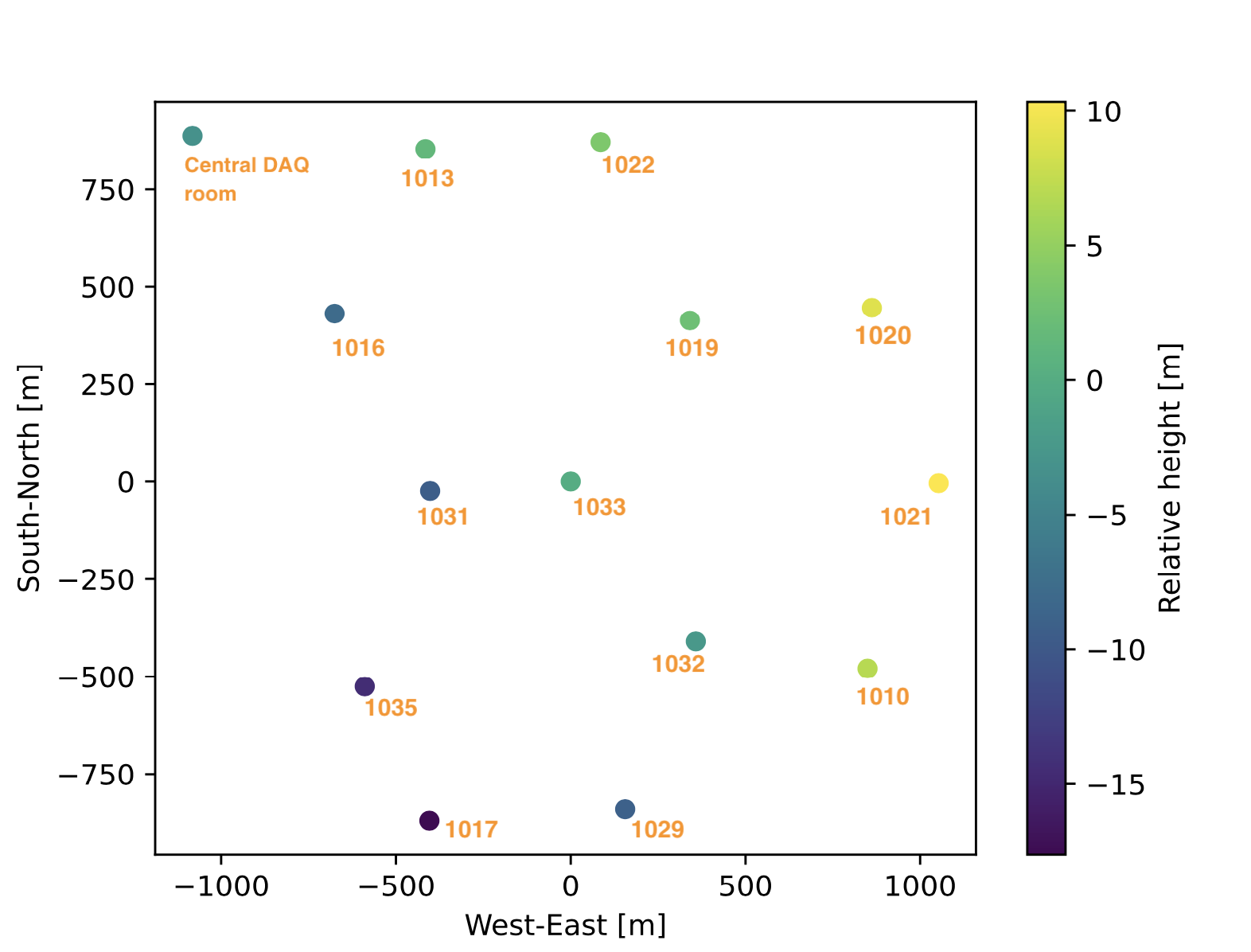}
\includegraphics[width=6cm]{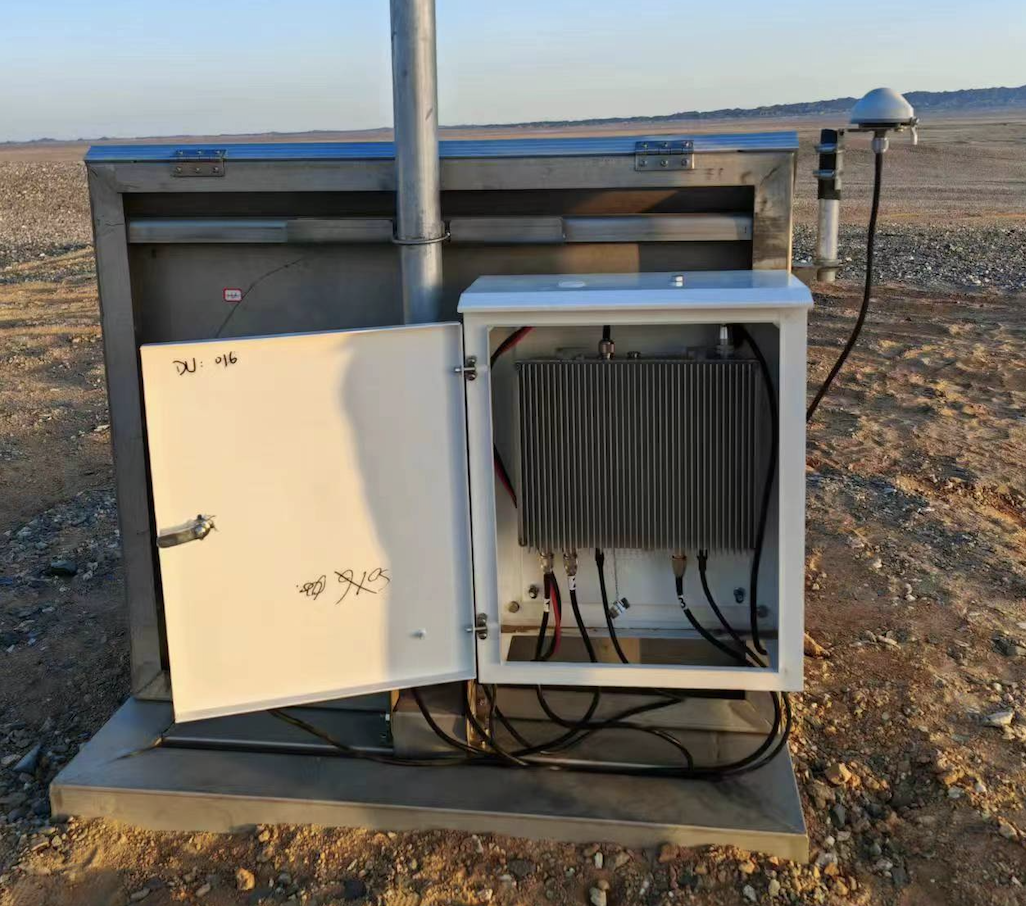}
\includegraphics[width=5cm]{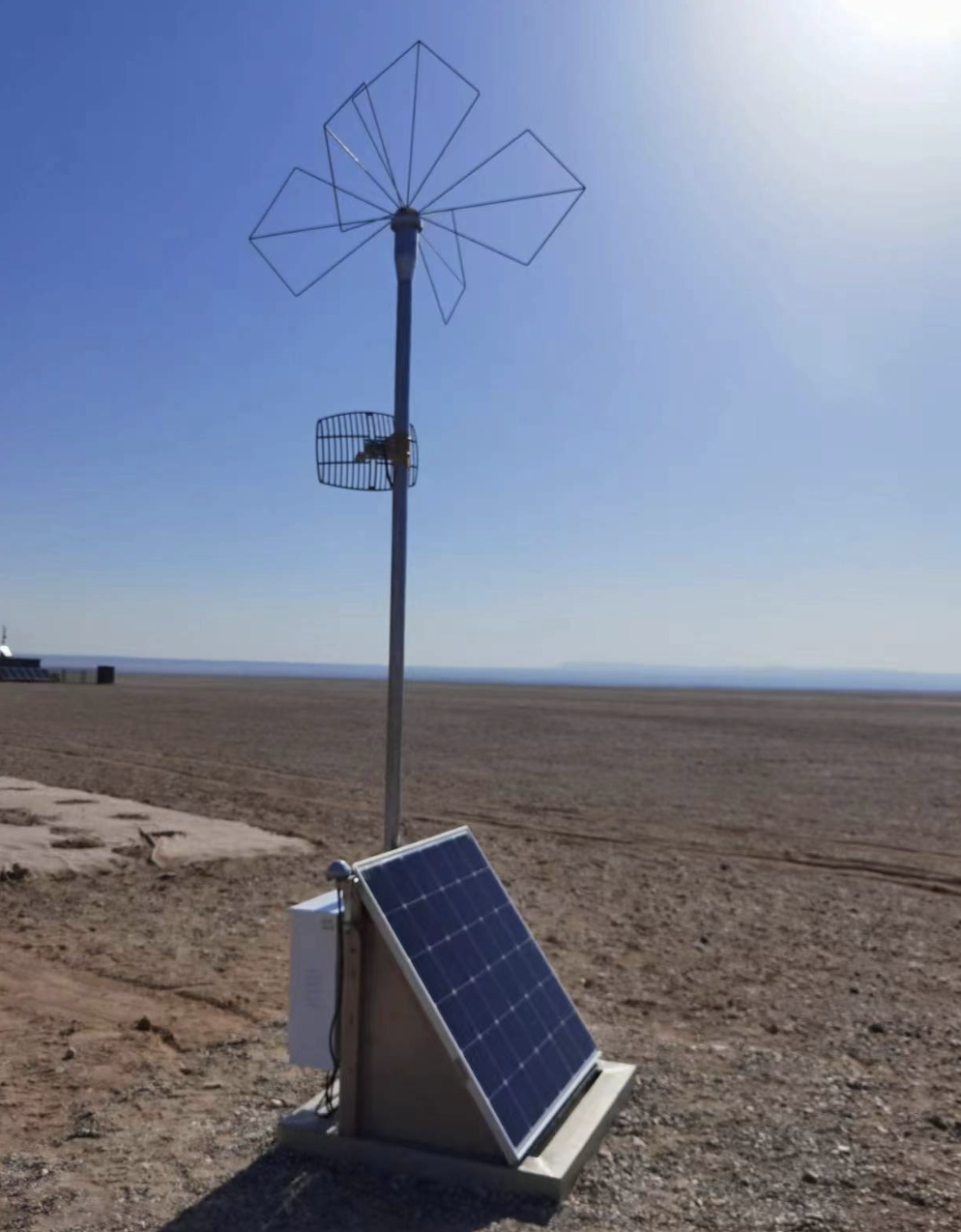}
\includegraphics[width=4.9cm]{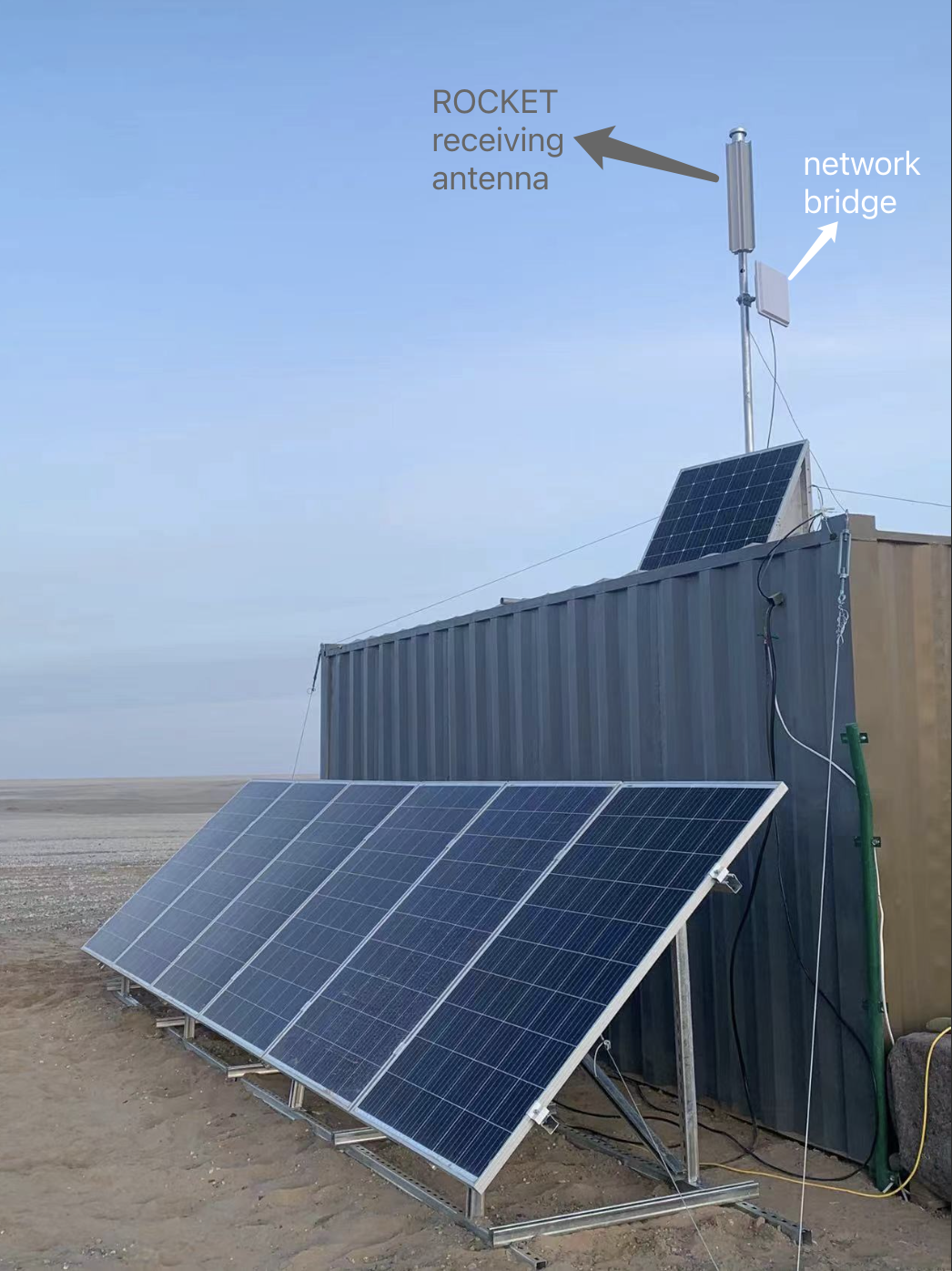}
\caption{The layout of GP13 at the Dunhuang site is shown in the top-left plot. The top-right and bottom-left photographs display one detection unit from the back and side views, respectively. In the top-left plot, the X and Y axes indicate the West-East and South-North directions, respectively. The color bar on this figure stands for the DU heights with respect to DU 1033, 1269 m above sea level. \mkred{The bottom-right plot shows the central DAQ room.}
\label{fig:layout}}
\end{figure*}

An array composed of 13 detection units (DUs),  called GRANDProto13 (GP13), has already been deployed in XiaoDuShan and is presently running to validate the DU design (see section \ref{sec:install}). A DU is composed of a so-called HorizonAntenna, a set of 3 bow-tie antennas designed
as depicted in the bottom plot of Fig.~\ref{fig:layout}. The three antenna signals are amplified by low-noise amplifiers (LNAs) plugged at the bow-tie antenna feed points, inside a mechanical structure called the antenna nut placed atop a 3.5-m pole. The signal is then transferred to a front-end board (FEB) at the foot of the pole, \mkblue{where the analog signal is amplified through a programmable
variable-gain amplifier (VGA), then filtered in the 30-230 MHz range}, and eventually digitized on 14 bits \mkred{by an analog-to-digital converter (ADC) with a sampling rate of}  500~\mkblue{MSamples per second}. \mkblue{The ADC conversion factor is $1~\mathrm{ADC} = 109.86~\mu \mathrm{V}$}. A system-on-Chip composed of a FPGA and 4 CPUs and placed on the FEB then deals with the trigger and event building. The DUs are also equipped with GPS antennas, \mkgreen{150-W} solar panels, and charge controllers for power supply. 

There are \mkgreen{a few} sensors mounted inside the nut at the top of the pole, which are used to measure the vibration of antenna, atmospheric temperature and pressure, and relative humidity. \mkred{In addition, a GPS sensor on the DAQ board allows us to measure the temperature inside the DAQ box.} Communication with the central DAQ system \mkred{is performed via WiFi} through a mesh antenna installed 2 m above the pole base as shown in the bottom-left panel of Fig.~\ref{fig:layout}. On the central DAQ side, an Ubiquity Networks ROCKET Prism RP-5AC-Gen2 installed at the roof acts as a receiver as depicted in the bottom-right panel of Fig.~\ref{fig:layout}. The detailed setup for each DU can be found in \cite{Zhang:2021tdh}. The distances between the central DAQ and each detection unit \mkred{of GP13} vary between 611 m and 2366 m. The GRAND DAQ system incorporates several operational modes to accommodate various scenarios. For the majority of GP13 data collection, a periodic mode with a 10-second interval is employed, meaning that data is sampled every 10 seconds. Additionally, the gains of the VGA in each detection channel are uniformly set to 20 dB.

\subsection{Installation}
\label{sec:install}

The installation of GP13 antennas took place between the end of February and the beginning of March 2023. A central data acquisition (DAQ) control room was established northwest of the GP13 array to conduct command transmission and data reception. \mkred{Internet access from the central DAQ room to the outside world is achieved with a network bridge at the roof pointing to a communication tower tens of kilometers away from the central DAQ room with download and upload speeds of 14 and 40 megabits per second, respectively. Solar panels supply \mkgreen{800 W} of power to the central DAQ room, which contains a computer connected to all 13 GP13 DUs.} \mkred{Since GRAND aims to detect inclined air showers,} the GP13 antennas were spaced approximately 500 meters apart, with some stations requiring relocation from their initial positions due to communication need, as depicted in top-left panel of Fig.~\ref{fig:layout}. After the installation in March, only five antennas were operational initially. This limitation was caused by the high temperatures experienced inside the DAQ box, \mkred{which at various times resulted in abnormal DAQ functioning and loss of connection}. To address this issue, a redesigned cooling system was successfully implemented for the DAQ, and \mkgreen{in May} all GP13 stations were equipped with new DAQ boxes at the back of the triangular structure hosting the battery and charge controller (\mkred{see top-right panel of Fig.~\ref{fig:layout}}). \mkblue{We implemented this \mkgreen{heat-dissipation} system using copper heat pipes that connect the chip and the box cover. Furthermore, we made modifications to the cover of the box, incorporating a heatsink to improve airflow and cooling efficiency.} Additionally, the LNAs were updated to \mkblue{significantly improve the impedance matching between the antenna and
the RF circuit.}

\begin{figure*}[!htb]
\centering
\includegraphics[width=7.5cm]{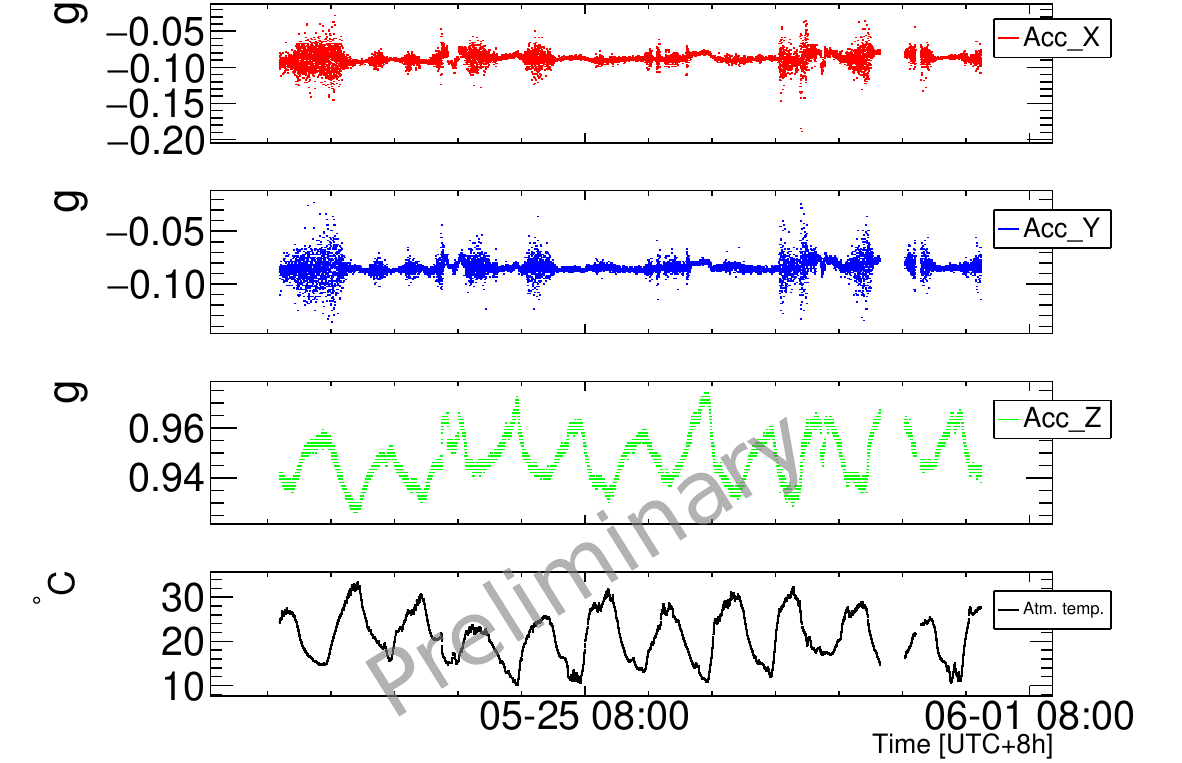}
\includegraphics[width=7.5cm]{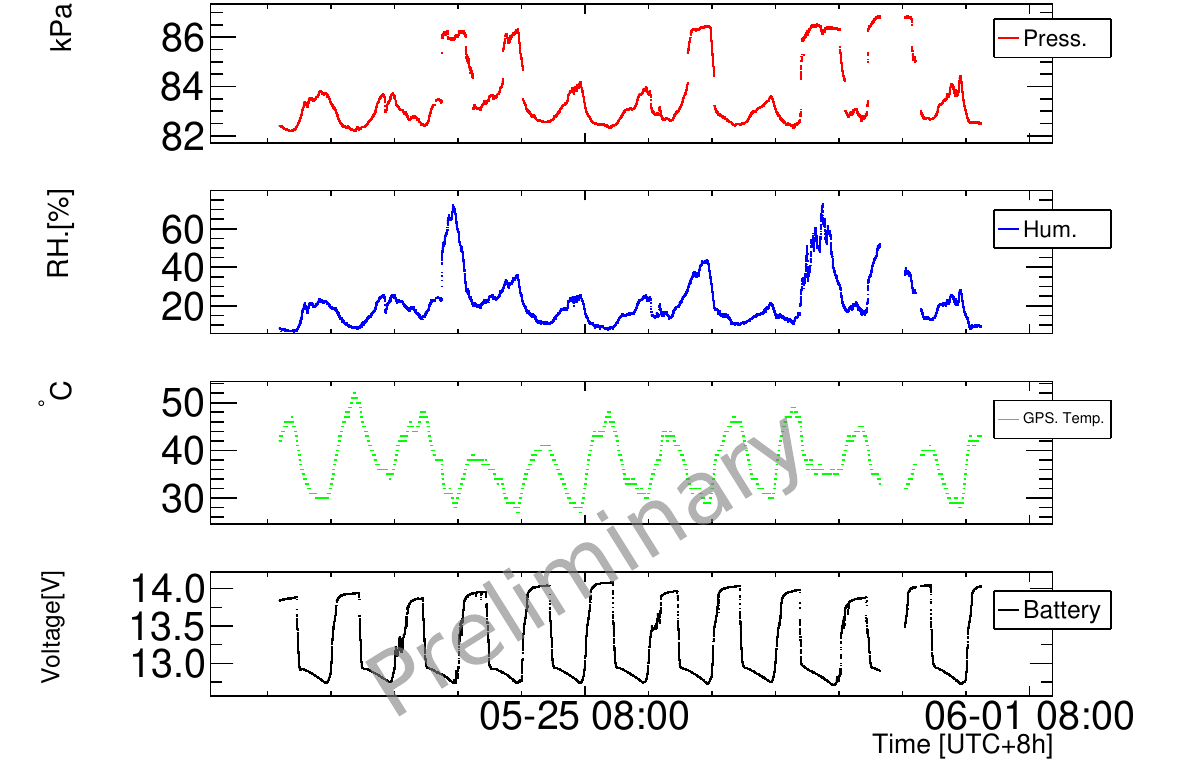}
\caption{\mkblue{The sensor parameters of DU~1010 over the last 11 days in May. From top to bottom, the left panel shows the acceleration in the X, Y, and Z directions in units of free-fall acceleration, $\mathrm{g}$, followed by the atmospheric temperature. The right panel displays the pressure, relative humidity, GPS temperature, and battery voltage level, from top to bottom.}
\label{fig:para}}
\end{figure*}

Starting from May 20, all 13 antennas of the GP13 array have been operational. Data were taken with the GP13 array from May 20 to May 31. Table~\ref{table:Trace} presents the data integrity for all 13 detection units over this period of time. "Expected." corresponds to the total number of traces that are supposed to be received during the run time, "Recorded." \mkred{corresponds to the} number actually recorded, and  "Valid." corresponds to the number of traces with valid length (i.e., 1024 samples). \mkred{We find that the number of recorded traces matches the expectations. The smaller number of traces for DU~1029 was induced by a temporary loss of connection with the central DAQ for a few hours.} We also found that non-valid traces correspond to an abnormal status of the FPGA register in the FEB for periods with high temperatures. Monitoring  parameters registered during this 11-day test period are also displayed in Fig.~\ref{fig:para}. The fluctuations of acceleration in both X and Y directions are strongly related to the speed of wind on site. \mkred{The non-zero baseline for both accelerations is thought to be a mis-calibration of the accelerometers. The measured acceleration in Z is correlated with the atmospheric pressure.} The significant variation of battery voltage reflects the charging process in daytime and energy consumption during night. \mkred{All of the sensors will be carefully calibrated in the near future.}

\begin{table}[!htb]
\begin{center}
\caption{Data integrity of the traces recorded in May for all GP13 detection units.}
\begin{tabular}{cccccccc}
\hline\hline
   DU ID & Expected & Recorded & Valid & DU ID & Expected & Measured & Valid  \\ \hline
   DU~1010 & 91224 & 91224 & 91207 & DU~1022 & 91224 & 91224 &  91210\\
   DU~1013 & 91225 & 91225 &  91212 & DU~1029 & 85416 & 85416 &  85405\\ 
   DU~1016 & 91224 & 91224 &  91210 & DU~1031 & 91225 & 91225 & 91212\\
   DU~1017 & 91225 & 91225 &  91215 & DU~1032 & 91224 & 91224 &  91212\\
   DU~1019 & 91223 & 91223 &  91211 &   DU~1033 & 91225 & 91225 &  91216\\
   DU~1020 & 91224 & 91224 &  91213 &  DU~1035 & 91224 & 91224 &  91213\\
   DU~1021 & 91224 & 91224 &  91214 & & & & \\

\hline \hline
\end{tabular}
\label{table:Trace}
\end{center}
\end{table}

\section{Data Analysis}

This contribution focuses on presenting the nighttime data collected during the final 11 days of May, \mkblue{specifically from around 22:00 to 05:00 in local time}.
 In the scope of this study, we have established a signal-to-noise ratio (SNR) metric to facilitate the automated search for pulse-like traces in Eq.~(\ref{equ.SNR}), without prior knowledge of the signal characteristics. The definition we employ for the SNR involves calculating the averaged root of the total squared value in the vicinity of the maximum ADC reading,
 divided by the root-mean-square (RMS) value of the part of the trace distant to the ADC maximum, depending on whether the maximum ADC value occurs early or late within the \mkgreen{traces}, 
\begin{equation}
\mathrm{SNR}=\frac{S - B}{B}.
\label{equ.SNR}
\end{equation}
\mkblue{Here, $S$ is the RMS value in the vicinity of the maximum ADC reading, while $B$ is the RMS value computed using the \mkgreen{remaining} segment of the trace preceding or following 300 ns to the maximum ADC reading. We defined a window \mkgreen{size} of $\pm$ 26 ns \mkblue{around the maximum} for the calculation of $S$. In case the ADC maximum is less than 26 ns away from the edge of a time trace, we define our window as [$-~N$, 26~ns] or [$-$~26~ns, $N$], where $N$ is the time from the maximum reading to the start or end of the trace, respectively.} \mkblue{Here $S$ is computed over a 52-ns integration window to mitigate the effect of traces with maxima associated with single-sample peaks.}

\begin{figure*}[!htb]
\centering
\includegraphics[width=7.5cm]{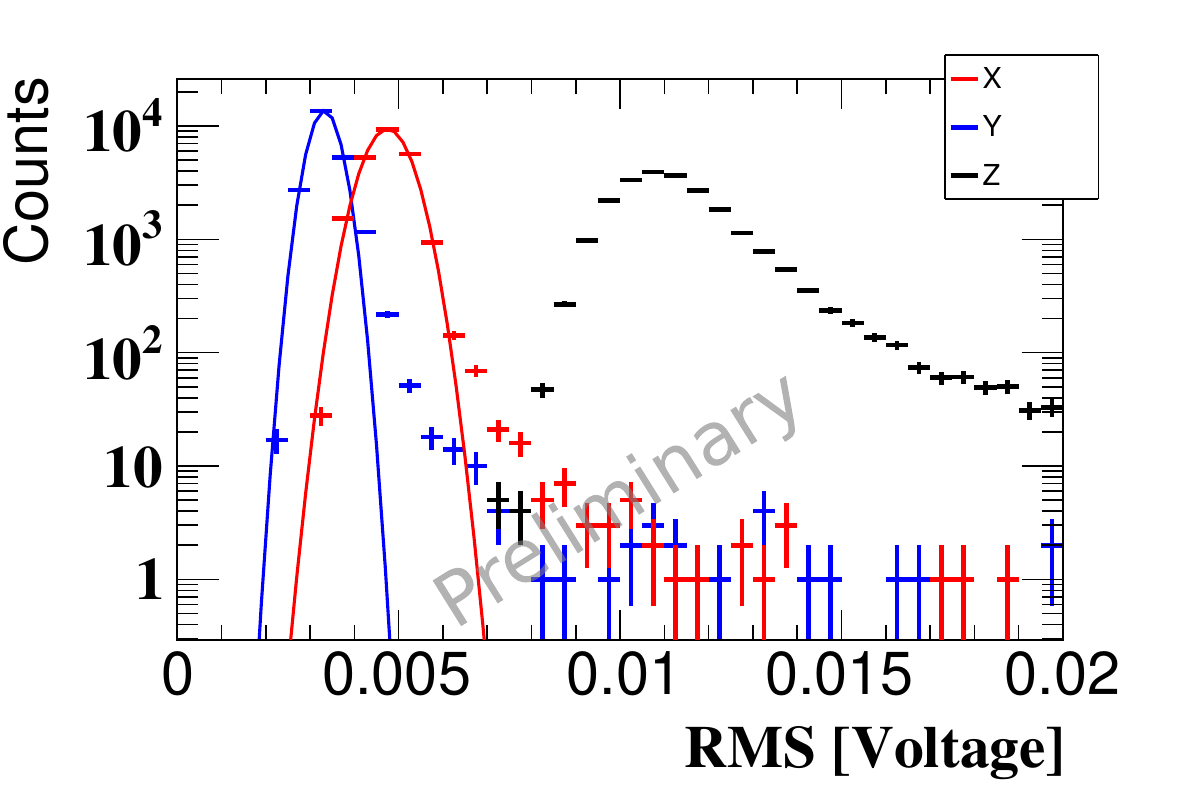}
\includegraphics[width=7.5cm]{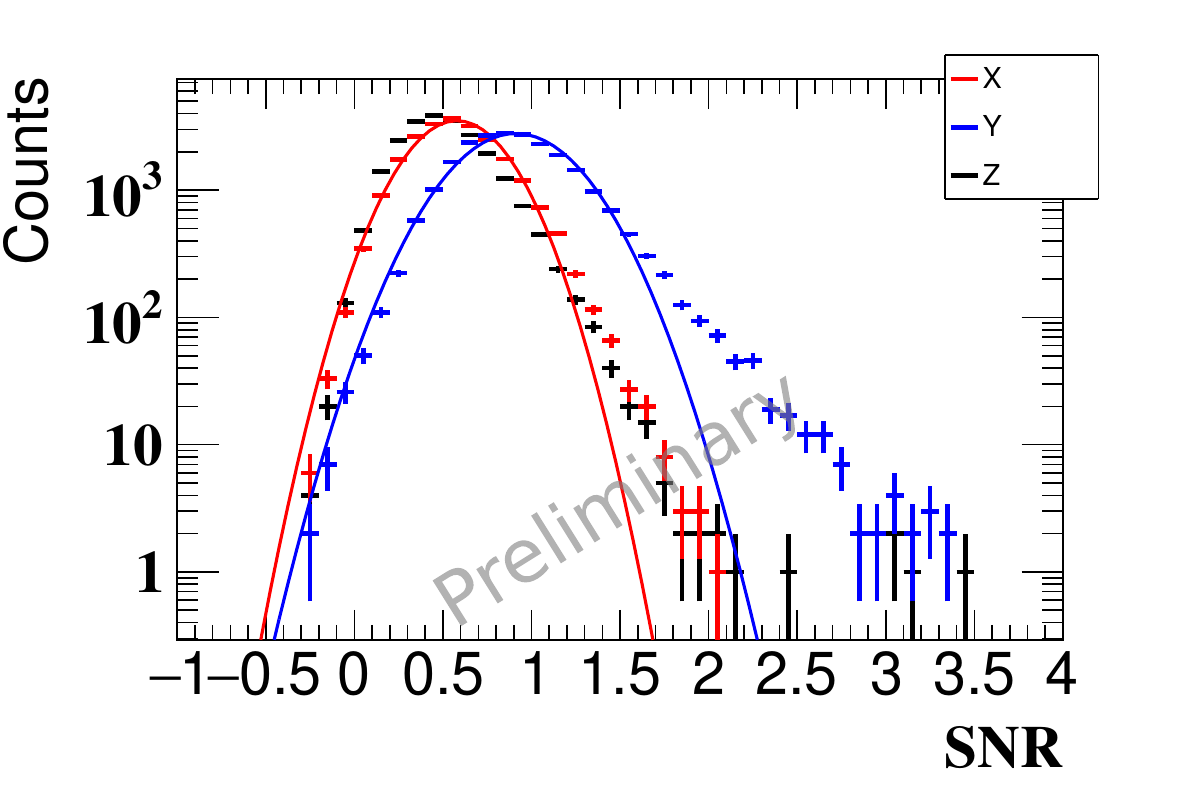}
\caption{The histograms of RMS (left) and SNR (right) of three channels from detection unit 1010 for all nighttime data during the last 11 days of May.  The red and blue lines in the two plots are Gaussian fits to the RMS and SNR of channels X and Y, respectively. 
\label{fig:RMS_SNR}}
\end{figure*}

Upon analyzing the nighttime data, we first filtered \mkred{the single frequency components coming} from civil aviation noise \mkgreen{in the 25--45 MHz and 115--145 MHz bands} \mkgreen{if there was a strong emission line in that channel compared to all other frequency bins}. The RMS distribution of the filtered traces is shown in the left panel of Fig.~\ref{fig:RMS_SNR}. For the X and Y channels, the low RMS values are well-described by a Gaussian distribution, with a pronounced tail at higher RMS values. Notably, the RMS values of the Z-channel exhibit the highest mean and the widest dispersion. In addition, \mkgreen{we applied a Gaussian fit to our SNR distribution for nighttime data (X and Y channels, see right panel of Fig.~\ref{fig:RMS_SNR}). Only a small fraction of data diverges from a Gaussian distribution. Our DUs captured traces with large SNR values, e.g., exceeding 2.5 in at least one detection channel. This illustrates the \mkgreen{relatively} low rate of transient radio signals observed at the XiaoDuShan site.}



\begin{figure*}[!htb]
\centering
\includegraphics[width=13cm]{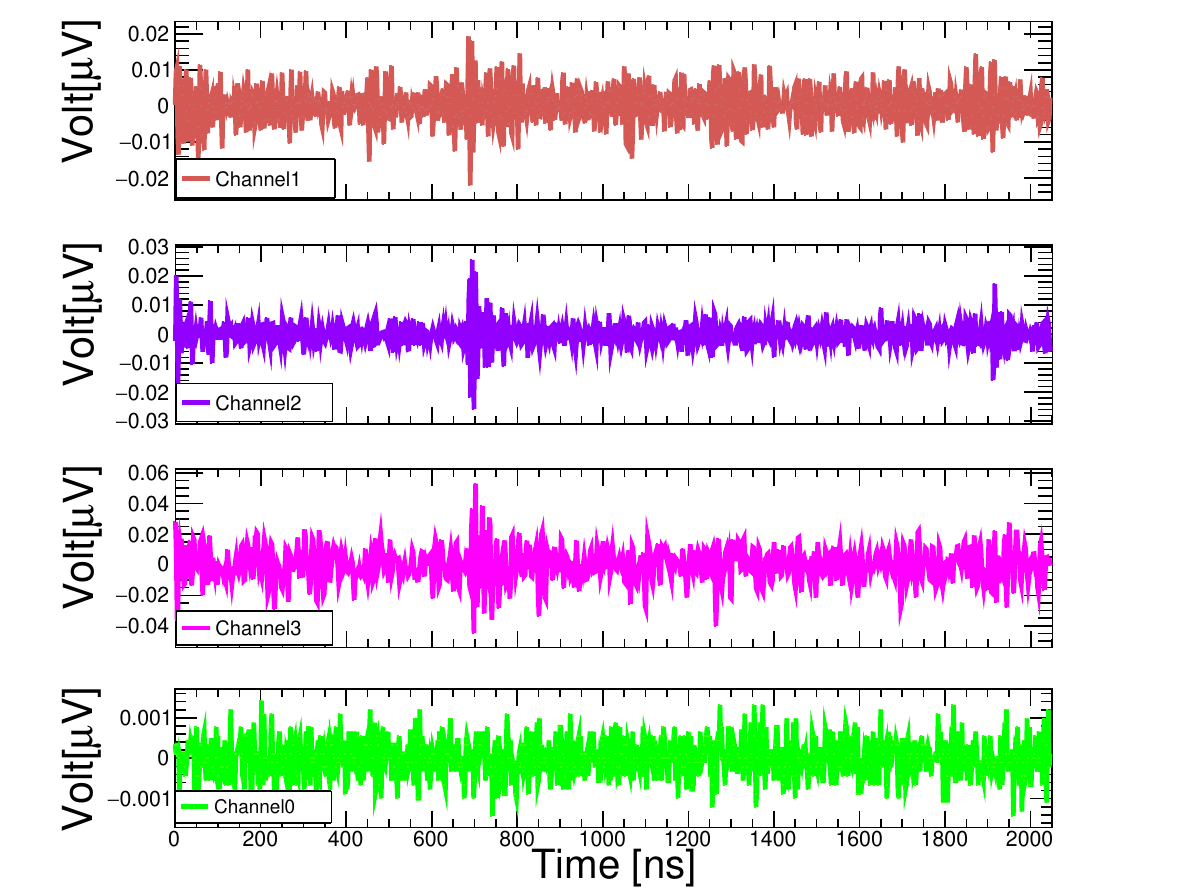}
\caption{One event taken by detection unit 1010 at the local time of 01:14 on May 21.  From top to bottom, each panel shows the measured time trace over sampling time of channel 1 (aligned with North-South, referred to as "X"), 2 (aligned with East-West, referred to as "Y"), 3 (aligned with vertical direction, referred to as "Z") and 0 (without channel connection), respectively.  
\label{fig:event}}
\end{figure*}

\begin{figure*}[!htb]
\centering
\includegraphics[width=16cm]{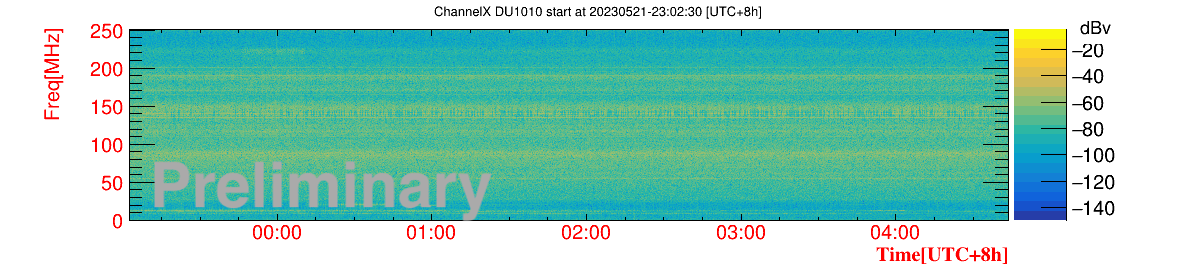}
\includegraphics[width=16cm]{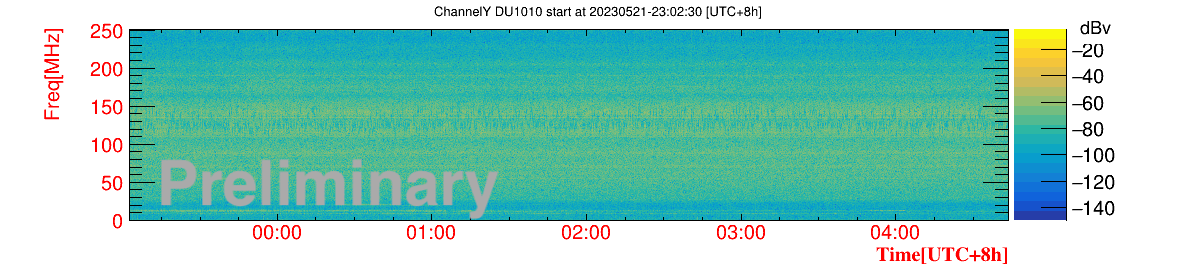}
\includegraphics[width=16cm]{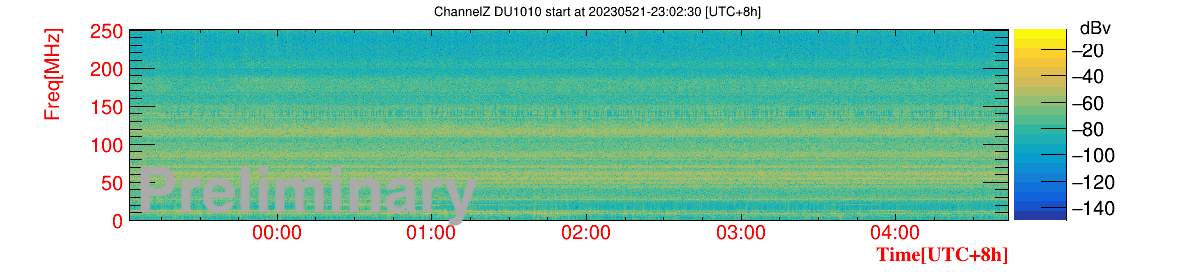}
\caption{The frequency spectrum of detection unit 1010 over a period of 6 hours during the night of May 21-22, 2023. The rows in the figure represent Channels X, Y, and Z from top to bottom, respectively. The spectrum is given in dBv units.
\label{fig:time}}
\end{figure*}

To quantify the strength of the noise, we performed a fast Fourier transform on the \mkgreen{measured time traces}. As illustrated in each panel of Fig.~\ref{fig:time}, \mkblue{it is an extended spectrum over time. The color gradient indicates the strength of spectrum at each frequency bin}
The relatively smooth profile across the three channels, \mkblue{combined with the absence of SNR outliers in right panel of Fig.~\ref{fig:RMS_SNR},} indicates that the rate of transient noise pulses is relatively low during nighttime, and that there are no significant bursts of such transients pulses during the night. \mkgreen{Here, we want to show an example of traces with high $\mathrm{SNR}$, found in coincidence in X,Y channels. One potential pulse candidate was chosen from the nighttime data on May 21, as illustrated in Fig.~\ref{fig:event}. Here, the largest SNR (2.7) was recorded in channel 2, about 700 ns from the start of the trace, which was aligned with the west-east direction. However, we were unable to confirm it as a genuine cosmic-ray event at present due to the current operating mode lacking monitoring of coincident detection among a group of DUs.}

\section{Summary}

We have achieved a significant milestone in the development of the prototype array for GRAND. The installation of GRANDProto13, consisting of the first 13 detection units for the GRANDProto300 array at the XiaoDuShan site in northwest China, was completed successfully in late February 2023. Throughout this process, we encountered challenges related to overheating inside the DAQ box, which we were able to resolve by May. 

Our data analysis has provided us with two promising indications:


1. The current design of the detection unit has demonstrated its robustness during extended operation in extreme conditions.

\mkblue{2. The current system is running stably with 13 units now that heating problems are solved.}

 \mkred{We are now working at commissioning the present array and preparing the deployment of 70 additional GRANDProto300 detection units which have \mkgreen{already} been built.}

\section{Acknowledgements} 

This work is supported by National Natural Science Foundation of China under grants 12273114, 
the Program for Innovative Talents and Entrepreneur in Jiangsu. We thank the municipal government of Dunhuang for their support on experimental site survey and permission of installation.  

\bibliographystyle{apsrev}
\bibliography{refs}


%
%
%

\end{document}